# Discrimination Among Multiple Cutaneous and Proprioceptive Hand Percepts Evoked by Nerve Stimulation with Utah Slanted Electrode Arrays in Human Amputees

David M. Page, Suzanne M. Wendelken, Tyler S. Davis, David T. Kluger, Douglas T. Hutchinson, Jacob A. George and Gregory A. Clark

*Abstract*—*Objective:* **This paper aims to demonstrate functional discriminability among restored hand sensations with different locations, qualities, and intensities that are evoked by microelectrode stimulation of residual afferent fibers in human amputees.** *Methods:* **We implanted a Utah Slanted Electrode Array (USEA) in the median and ulnar residual arm nerves of three transradial amputees and delivered stimulation via different electrodes and at different frequencies to produce various locations, qualities, and intensities of sensation on the missing hand. Blind discrimination trials were performed to determine how well subjects could discriminate among these restored sensations.** *Results:* **Subjects discriminated among restored sensory percepts with varying cutaneous and proprioceptive locations, qualities, and intensities in blind trials, including discrimination among up to 10 different location-intensity combinations (15/30 successes, $p < 0.0005$). Variations in the site of stimulation within the nerve, via electrode selection, enabled discrimination among up to 5 locations and qualities (35/35 successes, $p < 0.0001$). Variations in the stimulation frequency enabled discrimination among 4 different intensities at the same location (13/20 successes, $p < 0.005$). One subject discriminated among simultaneous, alternating, and isolated stimulation of two different USEA electrodes, as may be desired during multi-sensor closed-loop prosthesis use (20/25 successes, $p < 0.001$).** *Conclusion:* **USEA stimulation enables encoding of a diversity of functionally discriminable sensations with different locations, qualities, and intensities.** *Significance:* **These percepts provide a potentially rich source of sensory feedback that may enhance performance and embodiment during multi-sensor, closed-loop prosthesis use.**

## I. INTRODUCTION

CLINICALLY-available arm prostheses do not currently provide amputees with naturalistic and detailed tactile sensory feedback. Sensation from a prosthesis has been shown to be important for performance of functional tasks and for prosthesis embodiment [1]–[4], and amputees indicate interest in having sensory feedback from their prosthesis [5]–[10]. Peripheral-nerve interface approaches, such as Utah Slanted Electrode Arrays (USEAs), cuff electrodes, transverse intrafascicular multichannel electrodes (TIMEs), flat interface nerve electrodes (FINEs), and longitudinal intrafascicular electrodes (LIFEs) have demonstrated the ability to evoke sensory percepts at different locations, and of different qualities (e.g., submodalities) and intensities on the missing hand of amputees [11]–[18]. However, few of these have formally assessed functional discriminability among sensory percepts at different locations, and of different intensities and qualities

[16], [18]. Basic functional discrimination has been shown for objects of different shapes/sizes and compliances during closed-loop prosthesis control [1], [4], [19], [20].

The USEA provides intrafascicular access to nerve fibers spanning the cross-section of a peripheral nerve via 100 penetrating microelectrodes. In contrast to other peripheral nerve interfaces, USEAs offer cross-sectional nerve access via many channels, enabling activation of numerous sensory percepts spanning the hand [15]–[17]. The selection of different stimulation electrodes enables activation of different axons or subsets of axons with different projected field locations on the hand, and potentially with different sensory qualities. The stimulus intensity at each location can be encoded based on the frequency of stimulation [15], [21]. Despite this understanding, prior publications using USEAs have not fully tested the extent to which human subjects can discriminate among multiple proprioceptive and cutaneous sensory percepts at different locations, and of different qualities and intensities, such as would be desirable during multisensor closed-loop prosthesis control.

The extent of restored sensory feedback needed to reproduce a naturalistic sense of touch via a prosthetic limb is uncertain. However, discrimination tasks in intact human hands suggest that encoding of many locations, qualities and intensities, potentially via different receptor subtypes, is likely needed. For example, cutaneous location discrimination in the intact hand has been performed previously via a 2-point discrimination task, in which functional discriminability was achievable for stimuli as close as 0.55 mm apart [22]. This high level of discriminability is likely attributable to intensity encoding via a population of afferents both close to, and distant from, the site of applied tactile pressure (receptor density is on the order of 1 per square millimeter on the palmar hand [23], [24]). Natural activation patterns in the human hand include activation of several different cutaneous mechanoreceptor subtypes innervating many different locations on the hand. The number of discriminable locations on an intact human hand has not been formally quantified, but, based on these prior publications, is likely on the order of hundreds of sensory locations.

In microneurography studies, intact subjects have also discriminated among tactile percepts with the same location, but with different intensities. A roughly linear, nearly 3-fold increase in perceived intensity was noted both for normal cutaneous forces between 1-5 N and tangential forces between 1-3 N [25], with an informal indication that subjects are likely capable of discriminating up to ~10 different constant-force levels within these ranges. Constant-force intensities are generally accepted as being primarily encoded in the firing rates and activation patterns of type I slowly-adapting receptors (e.g., Merkel disk receptors) [26]–[29], although many receptor subtypes are generally activated during naturalistic touch of an intact hand. Type I and type II rapidly-adapting cutaneous mechanoreceptors (i.e., Meissner and Pacinian corpuscles) are generally assumed to be the primary encoders of vibratory intensities via their population activation patterns and firing rates [26]. Human subjects have also been able to differentiate among at least 4 different amplitudes of vibratory tactile stimuli which were encoded with the amplitude (2.4 to 154 μm indentation) and frequency (10 to 200 Hz) of vibration [30].

Utah Electrode Arrays (UEAs) have been implanted in human somatosensory cortex and used to induce perception of tactile sensation in humans with spinal cord injury, including generation at different locations, and of different pressure levels [31]. Non-invasive approaches to restoring a sense of touch using cutaneous electrical stimulation have also been used [32]–[34]. However, these approaches do not provide direct access to the individual receptor subtypes and control of the population codes that are used in the intact hand.

In previous work, with four subjects referred to as S1-S4, we have shown that USEAs implanted in residual peripheral arm nerves of human amputees provide up to 131 sensations of various qualities and locations spanning the phantom hand of human amputees [16], [17]. However, past reports included only limited details regarding basic location and quality discrimination among cutaneous percepts for three subjects (S1-S3). Furthermore, previous reports did not include cutaneous intensity discrimination trials or discrimination among different proprioceptive digit positions, nor did they include discrimination trials for combinations of percepts with different locations and intensities such as would be presented during closed-loop prosthesis control.

In expansion of our prior work, we now provide additional results from three recent subjects, S3-S5, each of whom received implantation of two USEAs, one in the residual median nerve and one in the residual ulnar nerve. These results include discrimination among 5 or more cutaneous locations (S3 and S4), 4 levels of cutaneous pressure (S5), 7 proprioceptive digit-position combinations (S5), and 10 location-intensity combinations (S5). We also report delivering multielectrode USEA stimulation in a time-shifted, scheduled manner enables avoidance of current-summation effects. Avoidance of current summation allowed for simultaneous, multipercept sensation in subject S3 such as may be desired during multisensor, closed-loop prosthesis control.

## II. MATERIAL AND METHODS

### A. Volunteers

Three transradial amputees participated in this study, referred to as S3, S4, and S5. Subject S3 was a 50-year-old male with a left-arm amputation that had occurred 21 years prior. Subject S4 was a 36-year-old male with bilateral amputations that had occurred 16 years prior. Subject S5 was a 43-year-old male with bilateral amputations that had occurred 24 years prior. Each subject underwent psychological and medical assessments prior to participating in the study. Subjects were provided with training materials prior to implantation of the electrodes to allow them to learn the concepts and methods of the experiments in advance and thus reduce post-implant training time. These included mirror-box or prosthesis-video training materials [35] as reported with previous subjects [15]–[17]. The subjects were monitored for medical risks both during and after the implant period, and subjects S4 and S5 were treated for implant-related infections that resolved without issue. The consenting process and experimental procedures were approved by the University of Utah Institutional Review Board, and the Department of the Navy Human Research Protection Program.

### B. Device

Two USEAs (Blackrock Microsystems, Salt Lake City, UT, USA) were implanted in each subject: one in the median nerve and one in the ulnar nerve. The implant location for subject S3 was in the left forearm distal to the elbow, whereas the implants for subject S4 and S5 were placed midway along the left upper arm, proximal to the medial epicondyle and to many motor branch points. USEAs consisted of 100 silicon microelectrodes (sputtered iridium oxide) spaced 400 μm apart in a 10x10 grid across a 4x4 mm square base. The electrodes varied from ~0.75 to 1.5 mm in length to allow cross-sectional access to fibers at different depths within the peripheral arm nerves [36]. Separate looped platinum wires were also implanted as stimulation return leads and for use as recording reference and ground leads. These looped platinum wires were placed close to (within ~5 mm of) the USEAS at the time of implantation, and were generally sutured to the epineurium along with the USEA lead wires within a centimeter of each USEA [37]. Electrical connection to each USEA electrode was available via an external printed circuit board that was coupled percutaneously to USEAs via a bundle of gold lead wires. Connection of the external circuit board to stimulation and recording hardware was made via a ZIF-Clip-96 connector cable (Tucker-Davis Technologies Inc., Alachua, FL, USA) for S3 and S4, or a 96-channel Gator connector cable (Ripple LLC, Salt Lake City, UT, USA) for S5.

The slanted nature of the USEAs enables cross-sectional nerve access to fibers at different depths, thereby increasing the possibility of activation of different axons or subsets of axons with each electrode [36]. An effort was made during the implant surgery to implant USEAs into the nerves so that the electrodes were positioned squarely perpendicular to the length of the nerve, which maximizes the cross-sectional nerve coverage of the USEA electrodes. The two-dimensional distance between two electrodes on the cross-sectional projection plane is likely the most influential factor on their ability to activate different axons or subsets of axons (Fig. 1). The stimulation amplitude on a given electrode influences which axons near the tip of the electrode are activated, whereas the stimulation frequency influences their firing rate. The stimulation amplitude may also influence firing rate when modulated at peri-threshold levels, for example, when only a subset of stimulation pulses in a pulse train result in generation of an action potential [18].

### C.  Surgical and Experimental Procedures

Subjects were given prophylactic antibiotics the day before, the day of, and for several days following the implant surgery (100 mg minocycline, 7 days, twice per day). USEAs were implanted in each subject under general anesthesia, via similar methods to those described in past publications [37]. For subject S5, electromyography leads were also placed in the muscles of the forearm for recording purposes, as described in [20], [37]. After exposure of each nerve implant site, the epineurium was dissected away, and USEAs were inserted into the nerve using a pneumatic insertion tool (Blackrock Microsystems, Salt Lake City, UT, USA) [38]. USEA lead wires and reference and ground wires were sutured to the epineurium, and a collagen wrap (AxoGen Inc., Alachua, FL, USA) was secured around the USEA, nerve, and reference and ground wires using vascular clips (Fig. 2a). For subject S5, the epineurium was sutured around the USEAs and reference and ground wires prior to placement of the collagen wrap. Dexamethasone (0.1 mg/kg) was delivered intravenously to the subjects during surgical closure as a potential means for decreasing the foreign body response [39], [40].

The site of percutaneous wire passage (Fig. 2b) was redressed roughly once per week using an antibiotic wound patch (Biopatch, Ethicon US LLC, Somerville, NJ, USA). Subjects S4 and S5 both experienced infections at the USEA percutaneous wire passage site with subsequent full recoveries after USEA extraction and antibiotic treatment. Implants were removed after 4 weeks, 5 weeks, and 13 weeks, for S3, S4, and S5, respectively. The USEAs from subject S3 were removed along with the section of implanted neural tissue for histological analysis [41].

Experimental sessions were typically carried out 1-3 days per week, for 1-5 hours each. In addition to the stimulation-evoked sensory percepts reported here, experiments consisted of impedance testing, decoding of neuronal and myoelectric signals for prosthesis movement control, and closed-loop control of a prosthetic hand.

### D.  Microstimulation

Electrical stimulation was delivered using the IZ2-128 System (Tucker-Davis Technologies Inc., Alachua, FL, USA) for S3 and S4, or the Grapevine System (Ripple LLC, Salt Lake City, UT, USA) for S4 and S5. Stimulation pulses were biphasic (cathodic first) with each phase typically having a duration of 200 μs (with a 100-μs interphase interval). Stimulation amplitudes and frequencies were kept below 120 μA and 500 Hz. These stimulation parameters were approved by the University of Utah IRB and have been used with USEAs in multiple subjects before without any considerable neural damage [41].

Subjects used either custom software to indicate the location, quality, and intensity or size of each USEA-evoked sensory percept on the image of a hand, or verbal descriptions. Subjects selected percept qualities from a list or created their own descriptors as necessary. Representations of percept locations and sizes, such as those shown in Figs. 3-7, were created based on the subjects' software markings as well as their verbal descriptions as appropriate.

Full-USEA threshold maps were collected periodically for each subject, as described previously [15]–[17]. These full-USEA maps provided a basis for selection of the electrodes used in the discrimination trials reported here. The electrodes chosen for a given discrimination task were typically selected based on outputs of a threshold mapping session performed within a week prior. Electrodes were chosen which provided distinct percepts with the desired location and/or quality. The temporal stability of sensory percepts or impedance was not generally used as a criteria for selection. During discrimination trials for subjects S3 and S4, a 200-ms train of stimulation was delivered at 200 Hz each time the subject or an experimenter pressed a button. For subject S5, three or four 500-ms trains of 100-Hz stimulation (unless noted otherwise, such as during intensity-encoding sessions) were delivered at a 50% duty cycle after the subject or the experimenters pressed a button. Subject S5 was typically instructed to determine the final percept intensity, quality, and location classification on the basis of the percept evoked by either the initial train in a trial or the final train in a trial for a given session, although practices varied depending on the session. Prior to discrimination trials, the activation threshold amplitude for each electrode (in μA) was determined by incrementally increasing the amplitude until the subject perceived a sensation.

### E.  Discrimination Trials and Data Analysis

Discrimination trials were performed by all three subjects during different stimulation sessions.  A stimulation session typically included mapping the percept locations, qualities, and intensities associated with several different USEA electrodes, and then down selecting to electrodes and stimulation frequencies which represented a subset of locations, qualities, intensities, or combinations for formal discrimination trials.

For location discrimination trials, electrodes were preferentially down-selected to represent sensation on different gross anatomical hand regions, such as different digits and the palm. When available, electrodes that evoked sensations at the same location but with different qualities were used for quality discrimination trials. Intensity discrimination trials were performed using different frequency encoding patterns on the same electrode or subset of electrodes. The number of electrodes selected for each discrimination study was based on the time remaining in a session and the desire to compare multiple trials of stimulation on each electrode in a blinded, randomized fashion. No sensory discrimination tasks were attempted with more functional levels (e.g., location, quality, intensity, or combinations) than those reported herein. Discrimination trial results reported here were not pooled across sessions or subjects; however, we have included results from similar discrimination trial configurations for different subjects.

Discrimination experiments were performed by delivering randomly-ordered stimulation trials in which the subject was required to classify the location, quality, and/or intensity of the evoked percept for each trial. Stimulation conditions varied across trials, including stimulation via different USEA electrodes or combinations of electrodes, and/or use of different stimulation frequencies. Formal discrimination trials were preceded by informal practice trials in which the subject experienced each different stimulation condition and formulated category labels for the percept associated with the condition. Once the subject felt comfortable identifying the location, quality, and/or intensity of the different stimulation conditions, formal blind trials commenced in which the subject was required to select one of his predetermined percept categories in response to each stimulation trial.

For subject S3 and S4, discrimination trial stimulation conditions included different electrodes and combinations of electrodes. For subject S5, stimulation conditions included different electrodes and/or stimulation frequencies. Importantly, catch trials (no stimulation) were added as a stimulation condition for subjects S4 and S5 to test the hypothesis that sensory percepts were indeed evoked by USEA stimulation (in contrast to pseudesthesia). A summary of discrimination tasks and the number of USEA electrodes used can be found in Table 1.

Data analysis for discrimination trials was performed using the binomial test, where the probability of guessing the correct classification on a given trial was determined as the inverse of the number of predetermined classification categories. Hypothesis testing was performed with a critical value of $\alpha = 0.05$. A Bonferroni adjustment was made to the critical value for post-hoc tests by dividing the critical value by the number of post-hoc tests performed.

## III. RESULTS

All subjects successfully performed functional discrimination trials for percepts at different locations and of different qualities; percepts with the same location but different qualities; and percepts with the same location and quality but with different intensities. Additionally, subject S5 performed combined location/quality/intensity discrimination trials, including trial sets with cutaneous percepts and trial sets with proprioceptive percepts. Functional discrimination among percepts of different locations, qualities, and intensities will be important for future use of sensory feedback from multiple prosthesis-coupled sensors during closed-loop prosthesis control.

### A. Location Discrimination

Subject S3 successfully discriminated among 5 stimulation conditions that evoked sensation at five different hand locations: ring finger tip, little finger tip, little finger base, wrist, and combined perception at all four of these locations. These percepts were evoked by individual stimulation of four ulnar-nerve-USEA electrodes and combined simultaneous stimulation of all four of these electrodes, respectively. Stimulation amplitudes for the four electrodes ranged from 14-30 µA. The subject discriminated among these stimulation conditions by classifying the percept evoked into one of the 5 predetermined classification categories in 35/35 successful trials (20% chance per trial, $p < 0.0001$, binomial test; Fig. 3a). Importantly, the four electrodes selected for these stimulation trials had tip positions as close as ~899 µm within the nerve, yet they each evoked consistently unique sensory percepts, suggesting selectivity in axon activation. Additionally, the combined stimulation of all four electrodes did not result in emergent sensory percepts (a percept with different quality or location from the four individual percepts), suggesting that current summation during simultaneous stimulation was limited.

To better study current summation during simultaneous stimulation of multiple electrodes in subject S3, we selected two ulnar-nerve-USEA electrodes with tips placed less than ~899 µm apart within the nerve (~805 µm cross-sectional projection separation assuming USEAs were implanted squarely perpendicular to the nerve) and delivered four stimulation conditions: individual stimulation of each of the two electrodes in isolation, simultaneous stimulation of both electrodes with no time shift, and simultaneous stimulation of both electrodes with a 3-ms time shift relative to each other to produce an interleaved stimulation pattern between the two electrodes. Stimulation amplitudes for the two electrodes were 23 µA and 20 µA, and stimulation was delivered continuously for 4 s during each trial. The individual stimulation via two different electrodes consistently produced sensations of little-finger-tip sting and lateral-palm tingle, respectively, whereas interleaved stimulation of these electrodes (3 ms time shift difference, 200 Hz) consistently reproduced both of these percepts concurrently with no emergent sensations, and simultaneous stimulation (no time shift difference, 200 Hz) consistently produced both of these percepts concurrently accompanied by an emergent 'massage' feeling bridging between them. The participant successfully identified among these four sensations in 20/25 trials (25% chance per trial, $p < 0.001$, binomial test; Fig. 3b). One plausible explanation for the emergent massage feeling during simultaneous stimulation with no time shift difference is that multiple additional axons may have been activated due to spatiotemporal current summation from the two electrodes [36].

Subject S4 also performed location-discrimination trials, including discrimination among eight different cutaneous stimulation configurations: individual stimulation of each of 3 ulnar-nerve-USEA electrodes, simultaneous combined stimulation using each combination of subsets of 2 of these 3 electrodes, simultaneous combined stimulation using all 3 electrodes, and no stimulation. The participant successfully identified between these sensations in 11/24 trials (12.5% chance per trial, $p < 0.006$, binomial test, Fig. 3c). Stimulation amplitudes on the three electrodes ranged from 7-13 µA depending on the electrode. Single-electrode percepts included sensation of tingle on the ring finger, touch on the little finger and palm (sometimes with an associated sense of little-finger movement), and tingle on the outer edge of the little finger. The precise nature of combination percepts was not fully documented prior to beginning the formal trials; but informally, the subject indicated that they consisted of a combined sensation of the percepts evoked by the individual electrodes, potentially with fused projected fields or emergent sensations. Importantly, these trials also included a condition of "no stimulation," which was not included in testing with subject S3.

Subject S4 successfully identified when stimulation was delivered compared with when no stimulation was delivered in 24/24 trials (50% chance per trial, $p < 0.0001$, binomial test). This indicates that percepts were indeed evoked by USEA stimulation (in contrast to pseudesthesia).

### B. Quality Discrimination

Subject S3 successfully discriminated between two evoked percepts with the same location, but with two distinct qualities, produced via stimulation of two different ulnar-nerve-USEA electrodes (Fig. 4). The tips of these electrodes were separated by ~2.1 mm within the nerve (~578 μm cross-sectional projection separation assuming USEAs were implanted squarely perpendicular to the nerve). Stimulation amplitudes for the two electrodes were 11 μA and 12 μA. Prior to formal discrimination trials, the subject identified the percepts evoked by these two different electrodes as having identical intensities and locations near the ring-fingertip ("Right on, exact same space"), but differing qualities of vibration and tingle, respectively. In subsequent formal trials, the subject consistently discriminated between the percepts evoked by the two electrodes in 30/30 trials (50% chance per trial, $p < 0.0001$, binomial test). We hypothesize that the different qualities of sensations are due to having activated two different sensory afferent subtypes. This result suggests that subjects may be able to discriminate among activation of different afferent subtypes that have overlapping projected fields.

### C. Intensity Discrimination

Subject S5 successfully discriminated among 4 different cutaneous-percept intensities, encoded via stimulation with different frequencies on a single median-nerve-USEA electrode that evoked a sensation of tingle on all four fingertips, although the percept seemed to isolate to the middle-finger alone during later stimulation trials. The stimulation amplitude used during trials was 25 μA. During informal practice trials, the subject designated four intensity levels as "high," "medium," "light," or "nothing," corresponding to stimulation at 100 Hz, 70 Hz, 35 Hz or no stimulation, respectively. During subsequent formal trials, the subject correctly classified these percept intensities in 13/20 trials (25% chance per trial, $p < 0.005$, binomial test, Fig. 5).

### D. Combined Location and Intensity Discrimination

Subject S5 performed combined location- and intensity-discrimination trials, similar to what may be used as part of a multisensor closed-loop prosthesis. Trials were performed for both cutaneous and proprioceptive percepts, each with multiple intensity levels encoding either cutaneous pressure/touch, or joint position.

Three cutaneous percepts were encoded at distinct hand locations via median-nerve-USEA stimulation on three different electrodes, with associated percept descriptions of index-fingertip pressure, middle-fingertip touch, and palm pressure (stimulation amplitudes ranged from 17-64 μA). For each of these electrodes, stimulation was delivered at stimulation frequencies of 30 Hz, 70 Hz, or 100 Hz, corresponding to "light," "medium," and "heavy" touch or pressure for each location. Sham stimulation was also used (i.e., no stimulation), making a total of 10 classification categories (three intensities at each of three percept locations, plus sham). Subject S5 successfully discriminated among these 10 stimulation conditions in 15/30 trials (10% chance per trial, $p < 0.0005$, binomial test, Fig. 6). In post-hoc analyses, we found that most of the subject's success was attributed to accurate location discrimination; location was identified correctly in 26/30 trials (25% chance per trial, $p < 0.0005$, binomial test for location classification independent of intensity classification, using a corrected critical value of $\alpha = 0.005$). In contrast, intensity discrimination was successful but seemed challenging; intensity was identified correctly in 17/30 trials (25% chance per trial, $p = 0.02$, binomial test for intensity classification independent of location classification, using a corrected critical value of $\alpha = 0.005$).

Subject S5 also successfully performed combined location and quality discrimination for two proprioceptive percepts that encoded index-finger and middle-finger flexion positions, respectively, via median-nerve USEA stimulation (Fig. 7). Specifically, 17-μA stimulation was delivered at 30 Hz, 80 Hz, or 150 Hz on one median-nerve USEA electrode to encode 10°, 90°, or ~180°/fully-closed flexion on the middle finger (compared to rest position). On a different median-nerve USEA electrode (~1.6 mm away; ~409 μm separation in nerve cross-sectional projection assuming USEAs were implanted squarely perpendicular to the nerve), 40-μA stimulation was delivered at 100 Hz, 50 Hz, or 150 Hz to encode 20°, 50°, or ~180°/fully-closed flexion on the index finger. During practice trials, the subject felt strongly that the nonmonotonic frequency-intensity encoding for the index finger joint position was accurate. However, during formal trials, confusion among the 20°, 50°, and 180° conditions on the index finger was common. A sham condition was also included, creating a total of 7 classification categories (three intensities on each of two digits, plus sham). Subject S5 successfully discriminated among these proprioceptive digit and joint-position combinations in 21/40 trials (14.3% chance per trial, $p < 0.0001$, binomial test). The subject correctly identified which phantom digit moved in 32/40 trials (50% chance per trial, $p < 0.0001$, post-hoc binomial test for digit classification independent of joint-position classification, using a corrected critical value of $\alpha = 0.005$). The subject identified the correct joint position in 22/40 trials (33% chance per trial, $p < 0.005$, binomial test for joint-position classification independent of joint classification, using a corrected critical value of $\alpha = 0.005$).

### E. Qualitative Descriptions of Sensory Percepts

Our subjects reported enjoyment of the variety of sensations evoked by USEA stimulation, including both proprioceptive and cutaneous sensations. After his first stimulation session, subject S3 stated, "My hand is starting to stimulate like it's starting to wake up or something. It really feels good. […] It's good to know that there's something still there." In response to the proprioceptive percept of middle-finger flexion delivered during proprioception discrimination trials, subject S5 stated that the sensation felt "exactly like movement of the middle finger." When asked to describe one of the sensory percepts evoked during cutaneous location-intensity discrimination trials, subject S5 stated, "It feels like touch. It feels like if I touched that door." Subject S5 also indicated that depth perception was limited in the virtual environment.

## IV. DISCUSSION

We have demonstrated that USEA stimulation can be used to encode sensory percepts with functionally-discriminable locations, qualities, and intensities. Further, discrimination was possible for both cutaneous and proprioceptive percepts. Encoding of cutaneous sensory percepts with different locations and qualities was achieved by stimulation of different USEA electrodes or combinations of electrodes, presumably resulting in activation of different axons or subsets of axons within the nerve. Encoding of sensory percepts with different intensities was achieved by modulation of the stimulation frequency, presumably resulting in an increased firing rate in activated axons. We have also demonstrated that subjects can discriminate among multiple location-intensity combined percepts such as would be desired during closed-loop prosthesis control.

Additionally, we have shown that multi-electrode stimulation in an interleaved pattern allows for simultaneous activation of multiple sensory percepts without emergent sensations (in comparison with simultaneous stimulation patterns, which did evoke emergent sensations). Future use of simultaneous and interleaved multielectrode stimulation may allow for improvements in the number and nature of restored percepts. This result also provides an important proof-of-concept for a method of interleaving stimulation via different USEA electrodes when current-summation effects are not desired, for example, during closed-loop prosthesis control with simultaneous USEA-evoked sensory feedback from multiple prosthesis sensors. Although we have shown that USEA electrodes as close as 800 µm within the nerve cross-section can evoke distinct sensory percepts, simultaneous stimulation via these electrodes often results in current summation and potentially undesired activation of additional axons that evoke additional sensation. Use of interleaved stimulation allows for simultaneous generation of the individual sensory percepts without current-summation effects. During closed-loop prosthesis control, interaction with the external environment may result in simultaneous activation of multiple prosthesis sensors, potentially generating simultaneous stimulation via multiple USEA electrodes. Algorithms may be developed and incorporated to interleave stimulation on different USEA electrodes to prevent current-summation effects. One tradeoff of interleaving stimulation is that a more frequent occurrence of stimulation artifact will likely be produced in USEA electrode recordings, possibly interrupting the ability to perform neural recording decodes for prosthesis movement control. In this case, it may be desirable to develop stimulation artifact blanking approaches or to implant separate recording electrodes in a distant location where stimulation artifact will be minimized (e.g., the residual limb muscles or a distant nerve location).

It is uncertain in some cases whether proprioceptive percepts were elicited by activating proprioceptive fibers directly or by generating secondary proprioceptive signaling after activating motor fibers. In the majority of cases, we did not observe visible muscle twitching in the residual limb during USEA stimulation; however, there is a possibility that single motor fibers were activated without producing a visible muscle twitch. However, proprioceptive percepts associated with movements of missing hand muscles do not suffer from this potential confound and hence are likely due to direct activation of proprioceptive fibers.

Sensory feedback from the hand has been shown to be important for identifying when contact events between the hand and the environment occur and for identifying object properties such as curvature, texture, and weight. These complex properties are interpreted using sensory integration across various proprioceptive and cutaneous channels with many receptive fields. Cutaneous information, encoded via multiple different receptors (e.g., slowly-adapting I, slowly-adapting II, rapidly-adapting I, and rapidly-adapting II), provides information regarding contact locations, object texture, object slippage, and gross shape [26], [28], [42]–[45]. Proprioceptive channels provide information regarding hand conformation and position, which, in conjunction with cutaneous information, provides information regarding object shape, weight, and counterforce [46]. Many of these object properties are challenging to deduce using visual feedback alone, particularly when feedback is needed rapidly during motor tasks [47] or when handling opaque objects. The goal of functional discrimination among a variety of sensory channels is ultimately to provide the brain with sufficient information to deduce useful information regarding interactions with the external environment.

Our gross encoding of 3 stimulus locations, each with 3 different intensities, may be sufficient to assist subjects in identifying gross object properties such as size and compliance [4]. However, more complex properties such as curvature and skin indentation direction and force gradations will likely require encoding via sensory percepts of different submodalities (e.g., RAI and SAI) that have nearby projected fields [48]. Restored sensation via multiple axons with adjacent projected fields may be critical for naturalistic sensorimotor hand control because real-time neural encoding of object properties likely involves cortical comparison of spike timings from neurons with adjacent receptive fields [49]. We anticipate that functional prosthesis control will improve with increasing numbers and variety of discriminable sensory feedback channels. Importantly, the data reported in this paper represents the most complex sensory discrimination tasks attempted with subjects S4 and S5; future studies should be designed to attempt more sophisticated discrimination tasks.

In addition to functional performance benefits of discriminable, multi-sensor prosthesis feedback, we anticipate that there will be substantial psychological benefits to restoring sensory feedback to amputees, such as prosthesis embodiment and reduction in phantom pain [2], [3]. We hypothesize that the sense of prosthesis embodiment will increase as a function of the number of discriminable sensory percepts provided for feedback.

The ultimate goal of restored prosthesis sensation is not just to provide subjects with a useful tool, but also to provide subjects with a prosthesis that is perceived by subjects as a replacement hand. Although the results of this report do not begin to approximate the sophistication of an intact hand (hundreds of discriminable cutaneous locations, and ~10 discriminable force levels), this work represents a substantial improvement from commercially available prostheses with no sensory feedback. Specifically, we have demonstrated that different USEA-evoked percepts are discriminable from each other, including up to 3 gross-level hand regions such as different digits and the palm, each with 3 different intensities. Although this does not recreating the performance of intact human hands, this level of discriminability has been shown to result in substantial functional improvements in fragile object manipulation and haptic perception [1], [4], and performance may improve with long-term use [50]. We have also shown that this discriminability can be achieved across multiple trials in a discrimination task during a single session, demonstrating that percepts are not pseudesthesias.

Ongoing work should focus on discrimination among successively closer projected fields to identify minimum discriminable distances. Additionally, interleaved, multielectrode stimulation strategies may produce surround inhibition effects that could improve functional discrimination. Although USEAs offer the highest channel count of any peripheral nerve interface, the 100 channels likely will not provide the incredibly fine level of resolution that would be required to completely restore sensory hand function. Development of a neural interface that may provide such resolution remains as a substantial challenge to the field.

One minor limitation of this report is the lack of control of the intensity and/or quality of sensory percepts during location discrimination trials. Specifically, subjects described the different sensory percepts by indicating both their location and their quality, raising the possibility that the discrimination may not have been performed based on location alone. The same argument can be made for the intensity of the percept, as the intensity of the sensation at each location was not carefully noted and was not controlled. Future studies regarding location discrimination should attempt to control for the quality and intensity of the percepts.

We have also demonstrated in this report that selective activation of distinct axons or subsets of axons is possible using USEA electrodes as close as ~800 μm within the nerve. Stimulation amplitudes were between 7-64 μA for the trials reported here, which apparently allowed for focal activation of axons within the local area of an electrode tip without activating axons associated with electrodes ~800 μm away. Future testing should be performed using closer electrodes, such as neighboring electrodes that are ~400 μm apart, to see if selectivity is achievable. Additionally, we anticipate that selectivity will decrease primarily as a function of cross-sectional projection distance, suggesting that electrodes that are directly distal/proximal to each other are less likely to evoke selective sensory percepts due to the possibility that the same axon(s) will pass near each electrode tip. More data from electrodes with a variety of different cross-sectional projection distances is needed to perform such an assessment. Future USEA designs may use a steeper slant to allow for improved selectivity along distal-proximal rows.

The results reported here used generally frequencies above 100 Hz for location discrimination trials, as these frequencies tended to produce consistently distinct percepts. Prior work has shown that both frequency and amplitude can be used to modulate the perceived intensity of percepts [18], suggesting that location discrimination would be comparable across stimulation frequency and amplitude as long as the perceived intensity was similar. Nevertheless, alternative methods for encoding intensity in sensory percepts should also be investigated including the use of stimulation amplitude or activation of multiple neighboring electrodes. Subject S5 often indicated that perithreshold stimulation amplitudes evoked weak percepts compared with the stronger percepts evoked by suprathreshold amplitudes at the same stimulation frequency (comfort-level amplitude was typically 5-10 uA above threshold amplitude). We hypothesize that this intensity change is not due to recruitment of additional nerve fibers at these increasing amplitudes, but rather is due to the increased probability of evoking an action potential with each stimulation pulse at suprathreshold levels (compared with perithreshold levels), effectively increasing the firing frequency of the axon(s). Inherent in this hypothesis is the prediction that an increase in stimulation amplitude may encode increasing intensities until a saturation-point is reached (i.e., when each stimulation pulse produces a single action potential in the nerve fiber). Future intensity-encoding experiments using frequency modulation should use suprathreshold stimulation amplitudes rather than perithreshold amplitudes to decrease stochastic variability in frequency encoding at the axon level.

Although functional discrimination among sensory percepts provides an important metric for demonstrating that percepts are distinct, other assessments may provide additional information. For example, the results presented here do not provide an indication of the theoretical resolution of percept intensities or locations. Future experiments should include mapping of the just-noticeable-difference (JND) between percepts at different locations or of different intensities [18], [34]. JNDs can also be quantified for percepts of different intensities to indicate, for example, the minimum discriminable frequency differences for stimulation on an electrode. JNDs should be mapped at multiple frequency levels to provide a test of Weber's law, which predicts that the JND will scale linearly with stimulation frequency [51].

Informally, we observed habituation of some sensory percepts during intensity discrimination trials in subject S5. We did not explicitly study the habituation phenomenon extensively. However prior research has documented this phenomenon for neural stimulation [52], and habituation of responses evoked by natural sensory stimuli is ubiquitous [53], [54]. Informally, we found that in some cases there seemed to be less nominal habituation if the stimulation was delivered at an amplitude at least 150% of threshold, and if we allowed for ~30 s of rest between each trial. Despite this, the subject's performance discriminating among intensities typically declined slightly as trials continued, and the subject tended to underestimate the percept intensity in later trials compared with earlier trials in a session. The habituation of USEA-evoked sensory percepts should be studied and understood explicitly in future studies, particularly for longer-duration sensory prosthesis use. Intact subjects exhibit habituation in response to tactile stimulation of the skin [55]–[59]. Past studies indicate that the rate of nominal habituation varies as a function of stimulus frequency and inversely as a function of stimulus strength for neural interface stimulation [14], [52], suggesting that algorithms that use either frequency or charge per pulse to encode stimulus intensity may have different consequences across time, even though the two parameters may be functionally interchangeable for a given time point. We hypothesize that use of suprathreshold stimulation intensities, or addition of interpulse variability into stimulation trains (in contrast to constant-frequency stimulation), to produce more biomimetic stimulation patterns [4], may help reduce the effects of habituation.

Although the sensory percepts restored via USEA stimulation are generally stable within a 2-3 h session, the projected field location, quality, and intensity associated with each electrode often varies across sessions [17]. Due to this limitation, we did not attempt to repeat identical functional discrimination tasks in different sessions. This instability may be due to a number of factors, including micromechanical shifts of the USEA relative to nerve fibers, the developing foreign body response to implanted USEAs, or degradation or failure of USEA electrodes and/or wire bundles [60]. Ongoing improvements to USEAs should continue, with reliability and longevity as a high priority. Longer-duration implants may also result in improved stability. Additionally, novel stimulation strategies, such as multielectrode stimulation, may decrease the variability in population encoding due to microshifts of USEAs. Also, multielectrode population encoding using biomimetic, receptor-type-specific stimulation patterns may decrease between-session variability in which axons are recruited, while also potentially improving the discriminability and naturalism of some USEA-evoked sensory percepts. USEAs and intraneural electrodes, in contrast to epineural cuff electrodes, are capable of communicating with the peripheral nervous system on its own terms by independently activating subsets of different populations of specific receptor types with known projected fields in naturalistic, custom-tailored, tunable patterns.

Ultimately, we foresee development of a closed-loop prosthesis system with multiple discriminable sensory percepts coupled to sensors that span a physical prosthetic hand for use in activities of daily living. We anticipate that discriminable sensory feedback via a prosthesis will enhance motor control, particularly in scenarios where visual feedback is limited or undesired. Also, we anticipate that discriminable, multisensor feedback with variable intensity and tunable quality will enhance the level of embodiment of a prosthetic limb, helping amputees to feel as though their prosthesis is a replacement hand, in addition to being a useful tool. Sensory feedback during closed-loop control, and any associated limb embodiment, may also alleviate phantom pain and many of the psychological difficulties associated with losing a hand.

## V. CONCLUSION

We have shown that human amputees implanted with Utah Slanted Electrode Arrays in their residual peripheral arm nerves can discriminate among a variety of restored hand sensations in blind trials, including: a) percepts with different hand locations, b) percepts with different qualities, and c) percepts with different intensities. Additionally, we have demonstrated that one subject was able to discriminate among cutaneous or proprioceptive percepts with different combinations of location and intensity, such as may occur during functional prosthesis use with multiple graded sensors for feedback. Furthermore, we have presented a multielectrode stimulation strategy using interleaved stimulation, which may be useful for evoking multiple sensory percepts concurrently without the effects of current summation during closed-loop prosthesis control. Our subjects enjoyed most of the sensory percepts and appreciated feeling controlled sensation from their amputated hand. Future work should include investigation of functional discriminability using multielectrode biomimetic stimulation patterns, as well exploration of the limit of functional discriminability resolution with USEAs. We hypothesize that functionally-discriminable sensory percepts with different locations, qualities, and intensities, used during closed-loop prosthesis control, will enable enhanced embodiment and improvements in motor performance for prosthesis users.

### ACKNOWLEDGMENT

This work was sponsored by the Defense Advanced Research Projects Agency (DARPA) BTO Hand Proprioception and Touch Interfaces (HAPTIX) program under the auspices of Dr. Doug Weber through the Space and Naval Warfare Systems Center, Pacific Grant/Contract No. N66001-15-C-4017, and the DARPA MTO under the auspices of Dr. Jack Judy through the Space and Naval Warfare Systems Center, Pacific Grant/Contract No. N66001-12-C-4042. Additional funding was also provided by the National Institutes of Health (NIH NCATS Award No. 1ULTR001067). This paper constitutes a chapter of the doctoral dissertation of David M. Page, of the Department of Biomedical Engineering, University of Utah.

The authors further acknowledge the contributions of Richard A. Normann, Heather A. C. Wark, and Bradley Greger in prior studies related to this research. We also acknowledge Sliman J. Bensmaia and Hannes Saal for general consulting and input regarding the encoding of cutaneous sensation, and David

J. Warren for assistance in final revision of the manuscript and for general contributions to the design of hardware for these experiments.

**Table 1. Summary of Discrimination Tasks**

| Participant | Discrimination Task | Total Conditions | Number of Electrodes |
|---|---|---|---|
| S3 | Location | 5 | 4 |
| S3 | Location | 4 | 2 |
| S4 | Location | 8 | 3 |
| S4 | On-off | 2 | 1 |
| S3 | Quality | 2 | 2 |
| S5 | Intensity | 4 | 1 |
| S5 | Location-Intensity | 10 | 3 |
| S5 | Location-Intensity | 7 | 2 |

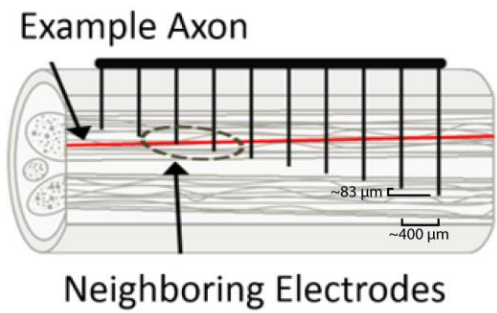

**Fig. 1. Absolute electrode distance versus cross-sectional projection distance. The 10x10 USEA provides cross-sectional coverage of peripheral nerves, increasing the possibility of activating different axons or subsets of axons with stimulation of each different electrode. Activation of different populations of axons is important for evoking sensory percepts with different locations or qualities. This diagram depicts a USEA implanted in a section of nerve, with an example axon that passes nearby two neighboring electrodes. Although the absolute distance between USEA electrodes is important for assessing stimulation selectivity limits, the cross-sectional distance between electrode tips more precisely indicates the likelihood that electrode tips are close to the same axon(s). For example, there is a ~409 μm absolute distance compared with a ~400 μm horizontal distance in and a ~83 μm vertical distance, not counting the exposure length of the electrode tip itself.**

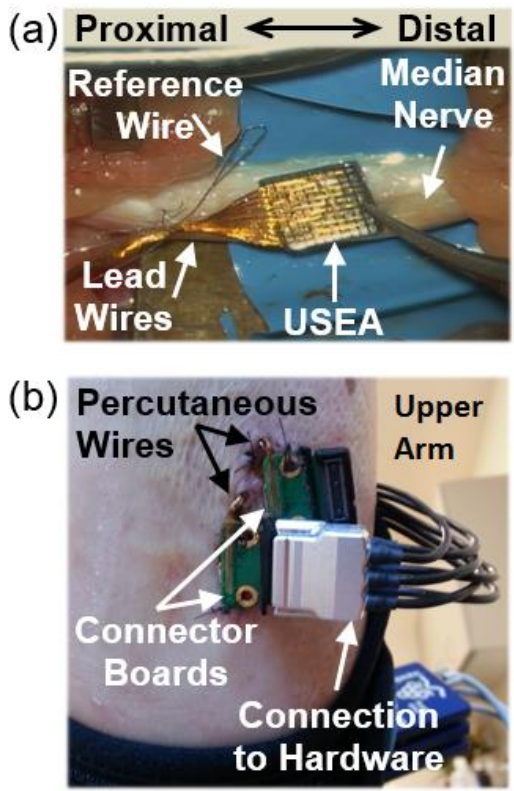

**Fig. 2. USEA implant methods. (a)** Photograph of a USEA in the median nerve of subject S4 taken shortly after pneumatic insertion. The bundle of gold lead wires as well as the separate ground and reference wires were later bundled to the nerve using a collagen nerve wrap. The USEAs were implanted with the long electrodes distally, to avoid damaging axons that may be recruited via stimulation of other USEA electrodes. **(b)** The USEA lead wires and ground and reference wires for each USEA (one in the median nerve; one in the ulnar nerve) remained attached to external connector boards via percutaneous incisions on either the lower or upper arm (subject S3 lower arm, subjects S4 and S5 upper arm). Stimulation hardware was attached to one or more of these external connectors during experimental sessions.

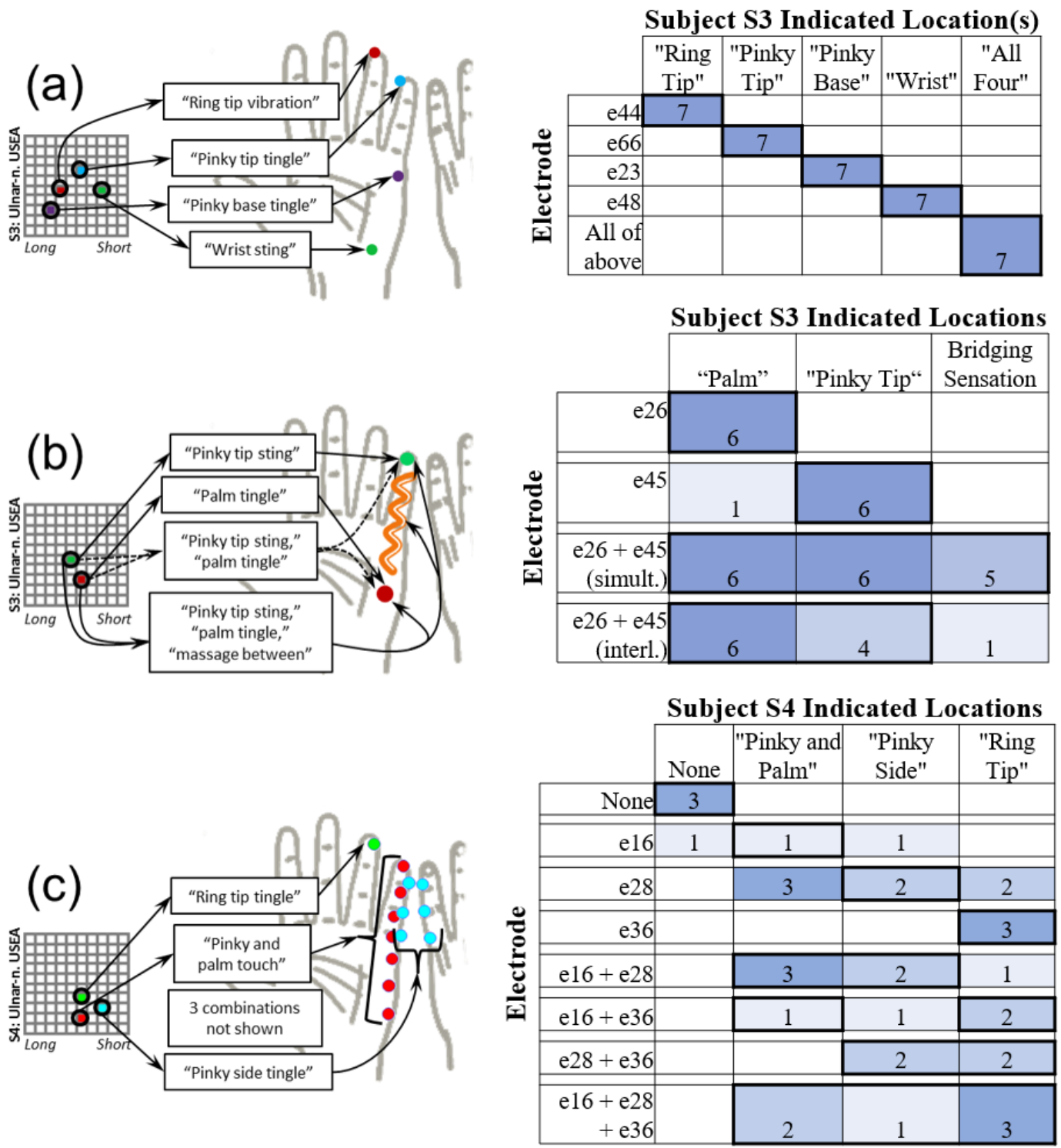

Fig. 3. Location discrimination trials. Each sub-figure depicts the location of electrodes which evoked different hand sensations, as well as a table showing the number of trials of the specified electrode or combination that evoked sensation in each hand location. For trials evoking multiple hand locations, the trial count in the cell for each designated location was incremented. Bold borders in tables indicate the 'correct' answer as designated in pre-trial programming. (a) Subject S3 successfully discriminated among percepts evoked via individual stimulation of 4 different ulnar-nerve-USEA electrodes (electrode locations on the USEA are shown as colored circles in figure), as well as simultaneous stimulation of all 4 electrodes (4 categories shown, 5th category was concurrent perception at all four locations; 35/35 successful trials, p < 0.0001, binomial test). (b) Subject S3 also discriminated successfully between simultaneous (converging solid arrows) versus interleaved stimulation (converging dashed arrows) of two ulnar-nerve-USEA electrodes, as well as individual stimulation of the two electrodes (separate solid arrows). Interleaved stimulation (3-ms time shift difference, 200 Hz) reproduced the original percepts simultaneously with no bridging sensation, whereas simultaneous stimulation (no time shift difference, 200 Hz) produced both of these percepts accompanied by an emergent "massage" bridging between them (20/25 successful trials, p < 0.001, binomial test). (c) Subject S4 discriminated among eight different stimulation configurations: individual stimulation of each of 3 ulnar-nerve-USEA electrodes, simultaneous combined stimulation using different subsets of 2 of these 3 electrodes, simultaneous combined stimulation using all 3 electrodes, and no stimulation (11/24 correct trials, p < 0.006, binomial test). Importantly, these trials with S4 also included a condition of "no stimulation", which was identified with 100% accuracy, indicating that percepts were indeed evoked by USEA stimulation (in contrast to pseudesthesia). These three experiments also demonstrate the selectivity of USEA-electrode stimulation, with unique percepts being generated by electrodes as close as 800 µm.

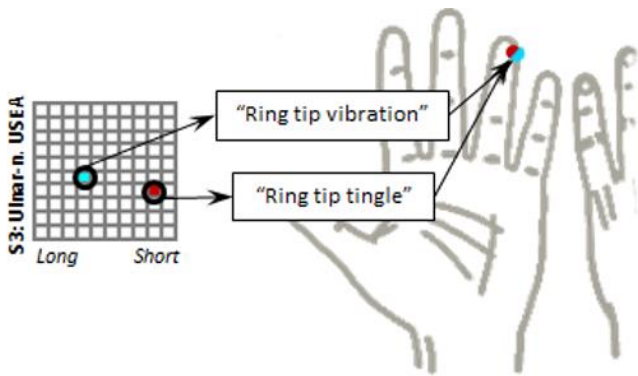

## Subject S3 Indicated Percept Quality

| | | "Tingle" | "Vibration" |
|---|---|---|---|
| **Electrode** | e39 | 15 | |
| | e44 | | 15 |

**Fig. 4. Quality discrimination trials. Subject S3 successfully discriminated between stimulation of two different USEA electrodes that evoked sensation at the same location, but with different qualities (vibration versus tingle). Regarding the locations of the two percepts, the subject said they were "Right on, exact same space." He also indicated that these sensory percepts were the same intensity level. The subject successfully performed the classification in 30/30 trials (p < 0.0001, binomial test). Electrode locations on the USEA are shown as colored circles in the figure. In the table, cell number/color indicates number of trials of the specified electrode that were assigned a perceived percept quality. Bold borders indicate the 'correct' answer as designated in pre-trial programming.**

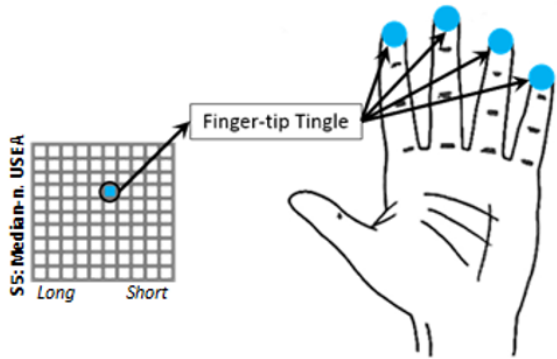

## Subjective Pressure Classification

| Frequency | | None | Light | Medium | Heavy |
|---|---|---|---|---|---|
| | 0 Hz | 5 | | | |
| | 35 Hz | 1 | 3 | | 1 |
| | 70 Hz | | 1 | 3 | 1 |
| | 100 Hz | | 3 | | 2 |

**Fig. 5. Intensity discrimination trials. Subject S5 discriminated between four percept intensities, evoked by stimulation of a single median-nerve-USEA electrode at three different frequencies (35 Hz, 70 Hz, 100 Hz) or sham (no stimulation). The evoked sensory percept was described as 'tingle' on all four fingertips, although in a later session this percept seemed to consolidate to the middle finger only. The subject successfully classified these different intensities in 13/20 trials (p < 0.005, binomial test). Electrode location on the USEA is shown as a colored circle in the figure. In the table, cell number/color indicates number of trials of the specified frequency that were assigned a perceived pressure level. Bold borders indicate the 'correct' answer as designated in pre-trial programming.**

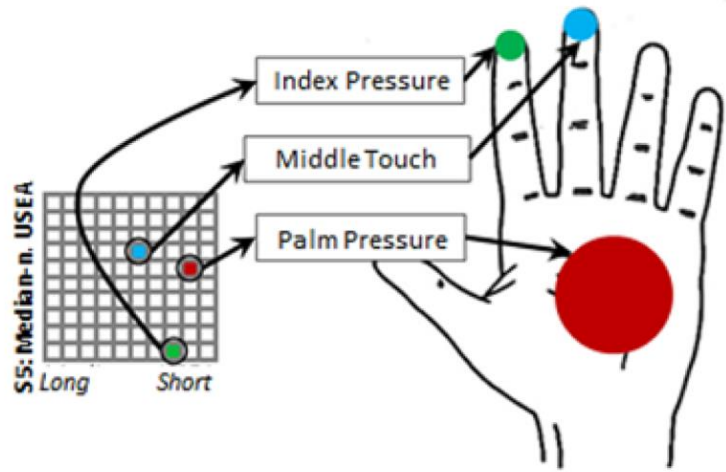

## Subjective Location/Intensity Classification

| | | None | Middle Finger | | | Palm | | | Index Finger | | |
|---|---|---|---|---|---|---|---|---|---|---|---|
| | | | Low | Medium | High | Low | Medium | High | Low | Medium | High |
| **Electrode/Frequency** | e66 — 0 Hz | 3 | | | | | | | | | |
| | e66 — 30 Hz | | 1 | 2 | | | | | | | |
| | e66 — 70 Hz | | | 2 | 1 | | | | | | |
| | e66 — 100 Hz | | | 2 | 1 | | | | | | |
| | e59 — 30 Hz | | 1 | | | 1 | 1 | | | | |
| | e59 — 70 Hz | | | | | 1 | 2 | | | | |
| | e59 — 100 Hz | | | | | | | 1 | | 2 | |
| | e8 — 30 Hz | | 1 | | | | | | 2 | | |
| | e8 — 70 Hz | | | | | | | | 1 | 1 | 1 |
| | e8 — 100 Hz | | | | | | | | 2 | | 1 |

**Fig. 6. Combined cutaneous location and intensity discrimination.** Subject S5 discriminated among combinations of different cutaneous percept locations and intensities. Three median-nerve-USEA electrodes evoked cutaneous "pressure" or "touch" percepts on the index finger, middle finger, and palm, respectively. Three frequencies (35 Hz, 70 Hz, and 100 Hz) were used to encode three different intensities via each electrode. Sham trials were also included (no stimulation) for a total of ten classification categories. The subject correctly classified the combination in 15/30 trials ($p < 0.0005$, binomial test). In post-hoc analysis, we found that most of the subject's success was attributed to accurate location discrimination (26/30 correct trials, $p < 0.0005$, binomial test for location classification independent of intensity classification, using a corrected critical value of $\alpha = 0.005$). Electrode locations on the USEA are shown as colored circles in the figure. In the table the cell number/color indicates the number of trials of the specified electrode (designated as e8, e569, and e66) and frequency that were assigned a perceived finger location and sensory intensity. Bold borders indicate the 'correct' answer as designated in pre-trial calibration programming.

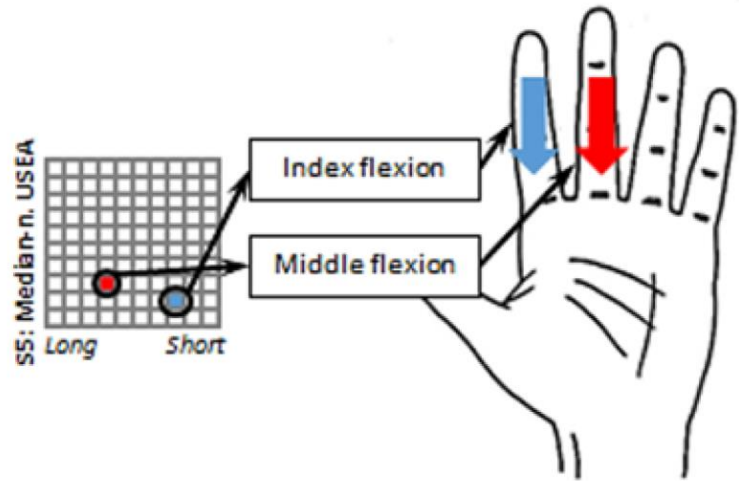

**Location/Intensity Classification**

| | | | Index Finger | | | Middle Finger | | |
|---|---|---|---|---|---|---|---|---|
| | | None | 20° | 50° | 180° | 10° | 90° | 180° |
| **e18** | 0 Hz | 8 | 1 | | | 1 | | |
| | 100 Hz | | 2 | | 3 | | | |
| | 50 Hz | | | 3 | 1 | 1 | | |
| | 150 Hz | | 2 | | 2 | | | 1 |
| **e24** | 30 Hz | | | 1 | | | 3 | 1 |
| | 80 Hz | | 1 | | 1 | 1 | 2 | |
| | 150 Hz | | | 1 | | | | 4 |

**Electrode/Frequency** (row axis label)

**Fig. 7. Combined proprioceptive location and quality discrimination.** Subject S5 discriminated between combinations of different proprioceptive percept locations and intensities. Two median-nerve-USEA electrodes evoked perception of proprioceptive flexion of the index finger and the middle finger, respectively. Three frequencies were used on each electrode to encode three different joint positions. Sham trials were included (no stimulation) representing a fully-open rest position for a total of seven classification categories. The subject correctly classified 21/40 trials (p < 0.0001, binomial test). Electrode locations on the USEA are shown as colored circles in the figure. In the table, cell number/color indicates the number of trials of the specified electrode (designated as e24 and e18) and frequency that were assigned a perceived finger and flexion angle. Note that the subject felt strongly during practice trials that the non-monotonic frequency-position encoding for the index finger was accurate, however we found that confusion between the 20°, 50°, and 180° conditions on the index finger was common during the formal trials. In the table, color shading scales with column percentage within each row. Bold borders indicate the 'correct' answer as designated in pre-trial calibration programming.